# MuBiNN: Multi-Level Binarized Recurrent Neural Network for EEG signal Classification

Seyed Ahmad Mirsalari*, Sima Sinaei+, Mostafa E. Salehi*, Masoud Daneshtalab+
*School of Electrical and Computer Engineering, University of Tehran, Tehran, Iran
+Division of Intelligent Future Technologies, Malardalen University, Vasteras, Sweden
{ahmad.mirsalari, mersali}@ut.ac.ir, {sima.sinaei, masoud.daneshtalab}@mdh.se

*Abstract*— Recurrent Neural Networks (RNN) are widely used for learning sequences in applications such as EEG classification. Complex RNNs could be hardly deployed on wearable devices due to their computation and memory-intensive processing patterns. Generally, reduction in precision leads much more efficiency and binarized RNNs are introduced as energy-efficient solutions. However, naive binarization methods lead to significant accuracy loss in EEG classification. In this paper, we propose a multi-level binarized LSTM, which significantly reduces computations whereas ensuring an accuracy pretty close to the full precision LSTM. Our method reduces the delay of the 3-bit LSTM cell operation 47× with less than 0.01% accuracy loss.

## I. INTRODUCTION

Monitoring the activities of human brain has great potential in understanding the functioning of our brain, as well as in preventing mental disorders and improve our quality of life. For this purpose, EEG systems have to be transformed from wired, stationary, and cumbersome systems often used in clinical practice today, to intelligent wearable, wireless, and comfortable lifestyle solutions that provide high signal quality. Continuous monitoring on wearable devices requires the automated EEG classification algorithm to be both accurate and light-weight at the same time. This forms our main focus in this paper.

Note that wearable devices have small and limited processors, which are much slower compared to desktop and server processors. Many previous algorithms are based on classical signal processing techniques [1][2]. Since the EEG signal characteristics significantly vary under different circumstances and for different people, the fixed features employed in such algorithms are not sufficient for accurately distinguishing among different types of disorders for all people. To extract the features automatically and increase the brain signal classification accuracy, deep learning based algorithms, including deep Convolutional Neural Networks (CNNs) and Recurrent Neural Networks (RNNs), have recently been proposed [3][4]. One of the most popular and effective RNN models used for sequence learning is called Long Short-Term Memory (LSTM) [5]. LSTM was designed to model the long-range dependencies, and memory backup of RNN plays a significant role, and so they are turned to be more accurate and effective than conventional RNNs.

This paper focuses on the EEG classification algorithm based on the LSTM recurrent neural networks. The proposed method employs RNNs because the EEG waveform is naturally fit to be processed by this type of neural network. RNNs capture the temporal dependencies in sequential data more efficiently compared to the other types of neural networks. However, high classification accuracy comes at high compute, storage, and memory bandwidth requirements, which makes their deployment particularly challenging, especially for resource-limited platforms such as portable devices. Compared to feed-forward Neural Networks, LSTM networks are especially challenging as they require state keeping in between processing steps.

Many techniques have been proposed to alleviate the compute and storage challenges described above. Among the most common ones are pruning [6], [7], [8], and quantization (see Section II for details) or a combination thereof [9]. All of them are based on the typically inherent redundancy contained within the NNs, meaning that the number of parameters and precision of operations can be significantly reduced without affecting accuracy [10]. Within this paper, we focus on the latter method, namely reducing computation cost and storage requirements through the reduction of precision in the leveraged data types. Ternary Neural Networks (TNNs) have attempted to reduce computation cost by considering three values ({-1,0,1}). TOT-Net as a remarkable TNN has deployed XOR and AND gates instead of multiplication [11]. The most severe quantization approach is binarization that results in Binarized Neural Networks (BNNs). BNNs have constrained operand values to {-1,1}, which eliminates most of MAC operations, and hence computations are remarkably reduced [14][16]. Therefore, they are extremely suitable for real-time and resource-limited embedded and wearable devices. However, deploying the conventional BNN model to datasets such as EEG, leads to remarkable accuracy loss. In this paper, many efforts are deployed to cope with the accuracy loss while still reduce computation complexity and memory storage.

In our previous work in [12], only the inputs and the weights of LSTM cell have been binarized and a set of coefficients in the form of powers of two have been considered for improving the accuracy and also it avoids computation-intensive multiplies. To the best of our knowledge, our proposed method called MuBiNN is the first effort to fully binarize LSTM for EEG classification with remarkable yield in complexity reduction. Our focus is on multi-level binarization of LSTM cells in each time step, for all weights, inputs, internal parameters, and outputs of the activation functions. Based on the binarization of all parameters, we propose an XNOR based multiplier for performing matrix and point-wise multiplications. MuBiNN significantly reduces computations and delays yet with an extremely negligible accuracy loss compared to the full-precision LSTM. The paper is structured as follows: in Section II, existing researches on quantized neural networks and their hardware implementations are reviewed, Section III presents a background on the LSTM. Our proposed method is discussed in Section IV. Section V presents the experimental results. Finally, the paper is concluded in Section VI.

## II. RELATED WORK

In Binarized Neural Networks (BNNs), the computations can be reduced with only binarized weights [13] or with binarizing both weights and activations. This method was firstly proposed in [14], [15]. Compared with the full-precision counterpart (32-bit), binarized weights drastically reduce memory storage as well as the memory accesses. On the other hand, BNNs significantly reduce the complexity of hardware by replacing costly arithmetic operations between real-value activations and weights with simple bit-wise XNOR and *popcount* operations, which altogether lead to much acceleration and a great decrease in chip area. In some applications, it has been shown that even 1-bit binarization can achieve reasonably good performance. In [16], a real-value scaling factor is proposed to compensate for the reduction of classification accuracy due to binarization and has achieved better performance compared to pure binarization in [14]. However, naïve binarization of both activations and weights leads to undesirable and nonacceptable accuracy reduction in some applications (like EEG signal classification) compared with the full-precision networks. To bridge this gap, recent works employ multi-level binarization with more bits like low bit-width networks [19], [20], [21], and ternary {-1,0,1} [17], [18] that achieve better performance and lead a useful trade-off between accuracy and implementation cost.

Among all existing researches on compression and quantization, most of them focus on CNNs, and less attention has been paid to RNNs. In fact, the quantization of RNNs has even more challenges. [19], [21], [22] have demonstrated high potential for RNN networks with quantized multi-bit activations and weights. Recently, [23] has shown that in some specific applications, LSTM cell with only binarized weights can even surpass the LSTM cell with full-precision weights. [24] has introduced a solution for quantizing both activation and weights in an LSTM network by formulating the quantization as an optimization problem. Using language models and image classification tasks, they have demonstrated that 3-bit quantization accuracy was equivalent to the original 32-bit floating-point model.

However, in more complicated applications such as EEG classification, our observations demonstrate that conventional binarized LSTM leads to considerable accuracy loss. Our proposed quantization approach is an extension to the state of the art LSTM binarization techniques. This approach not only reduces the computation-intensive MAC operations but also is suitable for applications in which the conventional binarized LSTMs lead to unacceptable accuracy. We substitute full-precision weights and activations with multi-level binarized weights and activations to significantly reduce computations and yet retain accuracy close to the full-precision LSTM.

## III. LSTM THEORY

For the sake of clarity, we review the basic LSTM approach, earlier presented in [5]. The LSTM architecture is composed of several recurrently connected "memory cells". An LSTM cell is shown in Figure 1. Each cell is composed of three multiplicative gating connections, namely input, forget, and output gates, and the function of each gate can be interpreted as write, reset, and read operations, concerning the internal cell state. The gates in a memory cell facilitate the keeping and accessing of the internal cell state over long periods of time.

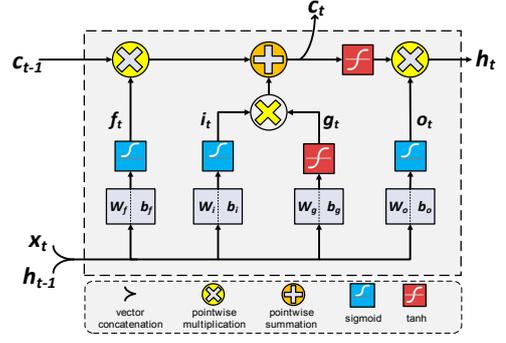

Figure 1. Long Short-Term Memory (LSTM) Cell.

Eq. (1) to (6) summarize formulas for LSTM network forward pass. The weight matrices are shown by $W_x$ and $W_h$. $x$ is the input vector, $b$ refers to bias vectors, and $t$ denotes the time (and so $t$-$1$ refers to the previous timestep). Activation functions are point-wise non-linear functions, that is *logistic sigmoid* for the gates and *hyperbolic tangent* for input to and output from the node.

$$f_t = sig(W_{f_x} x_t + W_{f_h} h_{t-1} + b_f) \quad (1)$$

$$i_t = sig(W_{i_x} x_t + W_{i_h} h_{t-1} + b_i) \quad (2)$$

$$g_t = tanh(W_{g_x} x_t + W_{g_h} h_{t-1} + b_g) \quad (3)$$

$$o_t = sig(W_{o_x} x_t + W_{o_h} h_{t-1} + b_o) \quad (4)$$

$$c_t = f_t \times c_{t-1} + i_t \times g_t \quad (5)$$

$$h_t = o_t \times \tanh(c_t) \quad (6)$$

According to the above equations, the LSTM output would depend on all previous inputs. Previous information is neither completely discarded nor completely carried over to the current state. Instead, the influence of the previous information on the current state is carefully controlled through the gate signals. In the full-precision LSTM, the length of all vectors is considered to be 32 bits.

## IV. PROPOSED BINARIZATION METHOD

As mentioned in Section III, LSTM has three distinct kinds of weights, including forwarding weight ($W_x$), recurrent weights ($W_h$), and bias weights ($b$). $x_t$, $h_{t-1}$, and $c_{t-1}$ are fed into the LSTM cell as inputs of the computing cell vector ($c_t$) and the hidden vector ($h_t$). In conventional LSTM, all inputs and parameters are full precision. Therefore, a considerable amount of memory and computation is required. For complicated applications like the EEG signal classification, LSTM networks become deeper/wider and hence more complex to achieve the required performance (inference time). On the other hand, wearable devices have limited computing resources, and the algorithm must run in an acceptable time frame. Therefore, continuous monitoring on wearable devices requires an accurate as well as light-weight EEG classification algorithm.

Although binarization techniques have been used in CNN [14][15][16], this work is the first effort to binarize LSTM with remarkable yield in accuracy while reducing the computation-intensive MAC operations. The proposed methods not only reduce the computation-intensive MAC operations but also are suitable for applications in which the conventional binarized LSTMs lead to unacceptable accuracy.

## A. The Naive Binarization method for LSTM (B_LSTM)

In binary neural networks, full precision parameters such as inputs (x) and weights (w) values are approximated with -1 or +1. The common method used for binarizing inputs and weights is called deterministic binarization functions according to (7) [14].

$$S_x = \begin{cases} +1 & if\ x \geq 0, \\ -1 & otherwise \end{cases} \quad (7)$$

Where $x$ is weight or input. Since the sign function is a deterministic function that has a simple hardware implementation, it is commonly used for input and weight binarization. The outputs of sign function {-1, +1} are encoded in binary {0, 1}, according to (8).

$$sign(x) = \begin{cases} +1\ (mapped\ to\ 1) & if\ x \geq 0, \\ -1\ (mapped\ to\ 0) & otherwise \end{cases} \quad (8)$$

By encoding the sign values $\{-1, +1\}$ to binary vectors $\{0, 1\}$, the dot product operations between input and weight can be computed by simple xnor- popcount operations [16] which is shown in Figure 2.

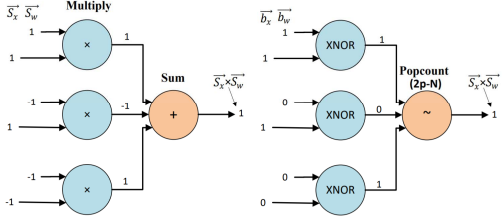

Figure 2. The relationship between the dot product (left) and xnor-popcount (right). Popcount operation computes 2p-N as result, where p and N denote the set bits and the length of the vector, respectively.

## B. The first proposed Multi-Level Binarization method for LSTM (MuBiNN1)

In the first method, we have proposed a specific multi-level binarization technique to quantize the inputs, weights, and internal parameters of an LSTM cell. The proposed method decreases the number of computations, resources, and memory footprint. Using a scaling factor for the binarized inputs and weights plays a critical role in decreasing accuracy loss in binarization. The proposed multi-level binarization is performed according to Algorithm 1.

| **Algorithm1:** Computing Multi-Level Binarization (MLB) |
|---|
| **Inputs**: $α_1, ..., α_M$ (scaling factors), x (Input of Multi-Level Binarization Algorithm including Weights and Activation) **Output**: $l_1, l_2, ..., l_n$ (Output of N-Level Binarization Algorithm) |
| 1: $r = x$ <br> 2: **for** i =1 **to** N **do** <br> 3:    $l_i \leftarrow Binarize((Sign(r))$ <br> 4:    $r \leftarrow r - Sign(r) \times α_i$ <br> 5: **end for** |

Where $N, x, l_i,$ and $r$ denote the number of binarization level, input of algorithm (that can be the LSTM input, weight, or internal parameters of LSTM cell), $i^{th}$ level of binarization, and the obtained error arisen from binarization, respectively. For computing $l_i$ in each level of the proposed method, the deterministic function Sign(x) is used. The first level of the algorithm takes x as input and approximates it by $α_1.sign(x)$, and it can be $\{+ α_1, - α_1\}$ based on the $sign(x)$. Then the error of binarization is calculated as the difference between x and the approximated value $\{+ α_1, - α_1\}$. The next level of binarization is approximated based on the error of the previous level of binarization. For an M-level binarization, this process is repeated M times. it should be mentioned that sign values $\{-1,+1\}$ must be encoded to 0 and 1, respectively. In M-level binarization, each layer needs M scaling factors. We have considered a distinguished scaling factor for each of the weights, biases, and inputs based on their values. Therefore, matrix multiplication between weights and inputs is approximated, according to (9).

$$x * w = α_x.MLB(x) * α_w.MLB(w) \quad (9)$$
$$= α_x * α_w.(MLB(x) * MLB(w)) = γ.(MLB(x) * MLB(w))$$

Where *x*, *w*, and *MLB()* denote inputs, weights, and our proposed method for multi-level binarization of inputs and weights. The scaling factors of inputs and weights are represented with $α_x$ and $α_w$, respectively. The scaling factors are constants because they have been primarily obtained statically during training. So, we can calculate γ in Equ. (9) offline and directly store γ instead of $α_x$ and $α_w$. It should be noted that the number of levels in input and weight can be different in the proposed method.

As mentioned before, in [12] to avoid computation-intensive multiplication and to exploit simple shift operations instead, the proposed scaling factors are in the form of power of two values such as 1, 1/2, 1/4, 1/8, etc. for input (x) and the related weight ($W_x$). In the first method (MuBiNN1), we have multi-level binarized the input matrix multiplications in Eq. (1) to (4) for all LSTM cells. On the other hand, by restricting the scaling factor to the powers of 2 values, multiplication can be replaced by the simple shift operation. We have explored the design space to find proper values for scaling factors and the appropriate powers of 2 scaling factor to reduce the accuracy loss. In[12], we have only multi-level binarized the input matrix multiplications in Eq. (1) to (4). This idea has significantly achieved better accuracies compared to the previous approach (B_LSTM), which are presented in the experimental results in Section V.

## C. The second proposed Multi-Level Binarization method for LSTM (MuBiNN2)

In the second proposed Multi-level Binarization method (MuBiNN2), we have binarized all parts of the LSTM Eq. (1) to (5). In order to achieve better approximation, the scaling factors are learned during the training phase. Each LSTM Layer has three separate kinds of weights, including forwarding weight ($W_x$), recurrent weights ($W_h$), and bias weights (b). The first one is multiplied to *x*, and the other one is multiplied to *h*. Also, each of them contains four separate weights ($W_g, W_f, W_i,$ and $W_o$) and four separate biases ($b_g, b_f, b_i,$ and $b_o$) as well. All these weights, internal parameters, and x as input are multi-level binarized by our method. Figure 3 has demonstrated the proposed architecture. The *xnor-based multiplier* unit is deployed to calculate the dot product results, and the obtained results are multiplied to the α value to achieve the final results. In Figure 3, $M_x, M_w, M_B, M_g,$ and $M_s$ denote multi-level binarized input, multi-level binarized weights, biases, cells, and scaling factors, which are stored in memory.

According to Figure 3, the proposed *xnor-based multiplier* module performs the matrix and vector arithmetic operations used to obtain the outputs of Eq. (1) to (4). The outputs of this component are passed to the *mult* unit, which is responsible for multiplying the scaling factor values to the output of the *xnor-based multiplier*. The output of this component is passed onto an *Add* component to add the bias to the result of the *mult* component. Then activation functions are performed. We choose piecewise linear approximation for the activation functions exploiting look-up tables. Next, xnor-based multiplier unit is deployed for performing the arithmetic operations of Eq. (5).

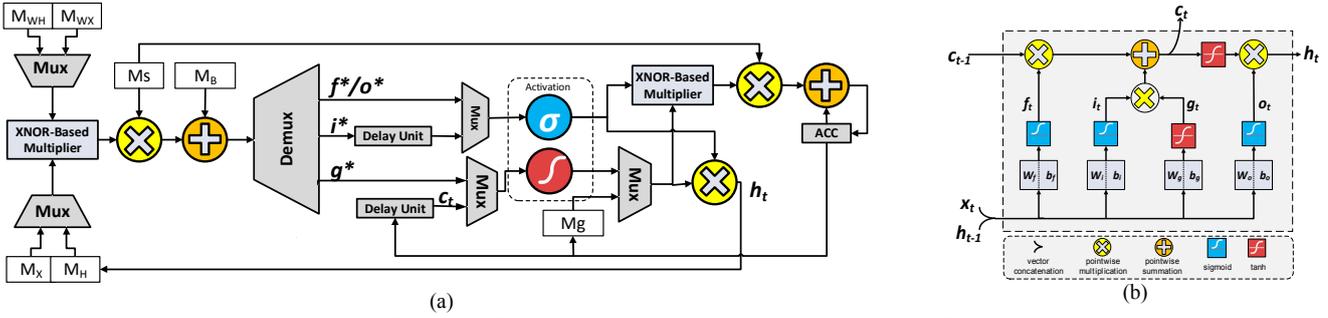

Figure 3. a) The proposed architecture unit for MuBiNN2, b) LSTM cell

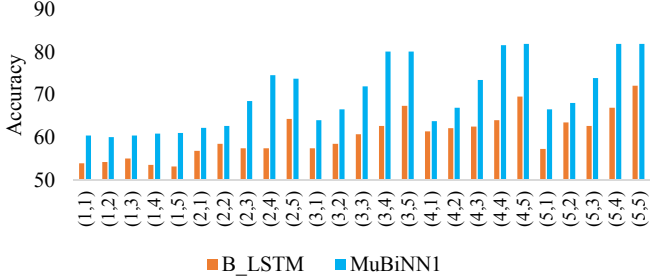

Figure 4. Comparison between the obtained accuracy of MuBiNN1 and B_LSTM approaches, (A, W) stand for the activation and weight bit levels, respectively.

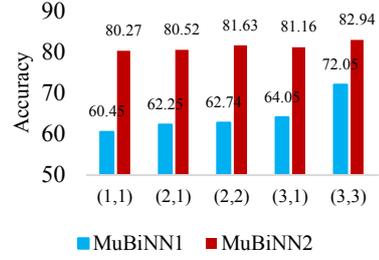

Figure 5. Comparison between the obtained accuracy of MuBiNN1 and MuBiNN2 method. (A, W) stand for the activation and weight bit levels, respectively.

Since the outputs of the activation function units are binarized, we should use the *xnor-based multiplier* to perform the Eq. (5). Finally, the MAC unit is used to perform the latest Equation in LSTM cell. In the second method (MuBiNN2), thanks to the learning of coefficients during the training phase, we were able to obtain better accuracy than the first method (MuBiNN1), as will be explained in the results.

## V. EXPERIMENTAL RESULTS

In this paper, we have binarized LSTM-based classification algorithm for EEG signal dataset. The conventional binarization method is not well deployed on EEG data set, and simple binary values lead to significant accuracy loss for this application. The number of features and time-steps of the deployed EEG dataset by 3625 samples are 32 and 1300, respectively. We have used Keras [25] library to implement our Software API. The network architecture consists of an LSTM layer with 100 output cells and a dense layer for final classification. The objective of our proposed method is to achieve the best accuracy. As mentioned in Section IV, conventional fixed-point binarization leads to remarkable accuracy loss, whereas our first proposed method (MuBiNN1) achieves close accuracy to full precision ones. Figure 4 has illustrated the best-obtained accuracies by our proposed B_LSTM and MuBiNN1 approaches. According to Figure 4, the MuBiNN1 method has significantly improved accuracy (more than 10%) compared to B_LSTM.

Figure 5 shows a comparison between the last two proposed methods (MuBiNN1 and MuBiNN2). It should be noted that the accuracy obtained in full precision is 83.67%. Actually, in our proposed method, based on the application constraints, the number of bit levels of weights and inputs can be chosen. We have reported the delay of the proposed architecture operations versus full precision operations. We have considered inference delay in terms of the logic depth and based on the gate delay values. Table I has demonstrated the delay of various combinations of (*Activation, Weight*) level binarized for MuBiNN2 and full precision inputs and weights. As shown in Table I, we have significantly reduced the delay of one (3,3) LSTM cell operations compared to full precision ones (~47×) while almost the same accuracy as the full precision accuracy has been achieved on the EEG classification by our proposed method.

TABLE I. THE LSTM CELL OPERATION DELAY IN SECONDS (IN 65NM TECHNOLOGY) FOR VARIOUS COMBINATIONS OF INPUT AND WEIGHTS BIT-WIDTH. FP STANDS FOR 32-BIT FULL-PRECISION.

| Activation /Weight | 1 | 2 | 3 | 4 | 5 | FP |
|---|---|---|---|---|---|---|
| 1 | 0.042 | 0.066 | 0.089 | 0.120 | 0.160 | 4.262 |
| 2 | 0.054 | 0.067 | 0.090 | 0.121 | 0.161 | 4.263 |
| 3 | 0.059 | 0.072 | 0.091 | 0.122 | 0.162 | 4.264 |
| 4 | 0.066 | 0.079 | 0.098 | 0.123 | 0.163 | 4.265 |
| 5 | 0.075 | 0.088 | 0.107 | 0.132 | 0.164 | 4.266 |
| FP | 1.065 | 1.079 | 1.098 | 1.123 | 1.154 | 4.294 |

## VI. CONCLUSION

In this paper, a novel multi-level binarized LSTM was proposed for the EEG classification algorithm. Our proposed method not only has significantly decreased computations in terms of delay, but also has almost achieved the accuracy close to full precision (83.67%) for multi-level binarization of activations and weights (80.27%, 81.63 and 82.94% by (1,1), (2,2), and (3,3) binarizations, respectively). The proposed technique is suitable for deployment on real-time and resource-limited embedded and wearable devices.

## ACKNOWLEDGEMENT

KKS has supported this work within the projects DeepMaker and DPAC.